\pdfoutput=1 
\documentclass[prl,twocolumn]{revtex4-1}

\usepackage{graphicx,color,transparent}
\usepackage{amssymb}
\usepackage{bm}
\usepackage{amsmath,amsfonts,latexsym}

\usepackage{braket} 

\usepackage{comment}

\usepackage{bm} 

\newcommand*{\cD}{{\cal D}}

\newcommand*{\cH}{{\cal H}}

\newcommand*{\cV}{{\cal V}}

\newcommand*{\aop}{a^{\vphantom{\dagger}}}
\newcommand*{\adop}{a^\dagger}

\DeclareMathOperator{\sgn}{\mathrm{sgn}}

\newcommand{\dd}[1]{\mathrm{d}#1\,}

\begin{document}

\title{The Fine Structure of the Phonon in One Dimension from Quantum Hydrodynamics}
\author{Tom Price}
\author{Austen Lamacraft}
\affiliation{TCM Group, Cavendish Laboratory, University of Cambridge, J. J. Thomson Ave., Cambridge CB3 0HE, UK}
\date{\today}
\email{tp294@cam.ac.uk}

\date{\today}

\begin{abstract}
We show that the resonant interactions between phonons in one dimension may be treated consistently within Quantum Hydrodynamics by the introduction of phonon dispersion. In this way the physics of a nonlinear Luttinger liquid may be described in terms of hydrodynamic (i.e. bosonized) variables without recourse to refermionization or the introduction of fictitious impurities. 

We focus on the calculation of the dynamic structure factor for a model with quadratic dispersion, which has the Benjamin--Ono equation of fluid dynamics as its equation of motion. We find singular behavior in the vicinity of upper and lower energetic thresholds corresponding to phonon and soliton branches of the classical theory, which may be benchmarked against known results for the Calogero--Sutherland model. \end{abstract}

\maketitle

One dimensional quantum fluids may be described within a hydrodynamic description usually known as Luttinger liquid theory~\cite{Haldane:1981}. This versatile framework has been applied to 1D gases of bosons and fermions as well as to spin chains and the chiral excitations at the edge of Quantum Hall fluids~\cite{Stone:1994,Giamarchi:2004,Gogolin:2004}. At the heart of the technique is the expression of all observables, as well as the Hamiltonian, in terms of bosonic collective variables describing the density and velocity, a procedure usually dubbed `bosonization'.

In recent years it has become clear that this approach suffers from serious shortcomings. Conventional bosonization treats phonons as linear excitations, described by a harmonic Hamiltonian, with no dispersion i.e. $\epsilon(k)=c|k|$, where $c$ is the speed of sound. Naively, one expects this to be a reasonable approximation as long as the anharmonicities present in a real system can be ignored. However, as we will make clear shortly, interactions between dispersionless phonons are singular in one dimension, and perturbation theory is inapplicable. As a result, a quantity as basic as the correct lineshape for the phonon excitations -- encoded in the dynamic structure factor -- appears beyond the reach of the usual theory.



Notwithstanding these difficulties, a `nonlinear Luttinger liquid' phenomenology has emerged in recent years, beginning with Ref.~\cite{Pustilnik:2006} and reviewed recently in Ref.~\cite{Imambekov:2012}. By a combination of methods based on refermionization, effective mobile impurity models, and exact solutions, the above difficulties have been sidestepped. However, these approaches hinge upon the introduction of degrees of freedom that are neither hydrodynamic nor microscopic, and whose existence can only be justified by appealing to continuity with weakly interacting or integrable limits. The fundamental conceptual question of how to describe the same physics within a theory of interacting phonons has hardly been addressed~\cite{Pereira:2007}. 

\begin{figure}[t]
\centering
    \includegraphics[width=\columnwidth]{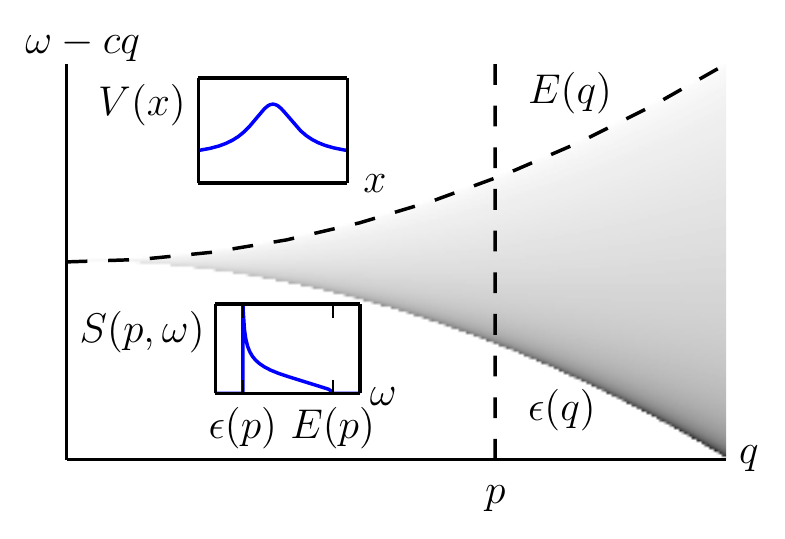}
    \caption{Dynamical structure factor $S(q,\omega)$ indicated by greyscale between the phonon $\epsilon(q)$ and soliton (dashed) branches $E(q)$, with  the power law behaviour of Eq.~\eqref{eq:Sbehaviour} as $\omega$ approaches the threshold at given momentum $p$ (lower inset). The upper inset shows a snapshot of the Lorentzian profile $V(x)$ of the soliton Eq.~\eqref{eq:1Soliton} in real space.}
	  \label{fig:Struct}
\end{figure}

In this Letter we provide a description of nonlinear Luttinger liquid physics solely in terms of Quantum Hydrodynamics, showing in particular how the dynamic structure factor acquires `fine structure' due to the nonlinearity. 
Our analysis hinges in an essential way on the inclusion of dispersive terms in the phonon Hamiltonian, in addition to the nonlinearity, which give rise at the classical level to two branches of excitations: small amplitude phonons and solitons (see Fig.~\ref{fig:Struct}). We show that the corresponding quantum theory yields predictions for the structure factor in agreement with the phenomenological nonlinear Luttinger liquid theory.

To illustrate the difficulties inherent in theories of non-dispersive phonons, consider the phonon Hamiltonian $H=H_2+H_3$ with $H_2 =\sum_{k>0} \epsilon(k)\, \adop_k\aop_k$, and the leading (cubic) nonlinearity with coupling $g$ \cite{Lifshitz:1980}
\begin{equation*}\label{eq:PhHam}
	\begin{split}
		H_3=\frac{g}{2} \frac{1}{\sqrt{L}}\sum_{k_{1}=k_{2}+k_{3}\atop k_{1},k_{2},k_{3}>0}\sqrt{k_{1}k_{2}k_{3}}\left(\aop_{k_{1}}\adop_{k_{2}}\adop_{k_{3}}+\text{h.c.}\right)
	\end{split}
\end{equation*}
%
Here $[\aop_p,\adop_{q}]=\delta_{p,q}$, $L$ is the system size, and we consider only right moving excitations with dispersion $\epsilon(k)=ck$, as it is interactions among phonons moving in the same direction that are resonant. The cubic terms in $H_3$ describe the disintegration of one phonon to two and the merging of two to one. By virtue of momentum conservation and the  linearity of the phonon spectrum, $H_3$ only couples states of the same energy, and is therefore a degenerate perturbation \cite{Samokhin:1998}. There is therefore no sense in which $H_3$ can be considered small. It is clear that this is a feature of \emph{any} interaction among linearly dispersing phonons moving in the same direction.

The same problem can be understood from a real space viewpoint by defining the usual chiral boson field
\begin{equation*}
	\label{OneLoopRG_expand}
	\phi(x) = -\sum_{k>0}\frac{i}{\sqrt{kL}}\left(\aop_{k}e^{ikx}-\adop_{k}e^{-ikx}\right),
\end{equation*}
with commutation relations $[\phi(x),\phi(y)]=\frac{i}{2}\sgn(x-y) $, in terms of which the phonon Hamiltonian takes the form
\begin{equation*}\label{eq:RealSpace}
	H = \int_{-\infty}^\infty dx\left[\frac{c}{2} \phi_x^2+\frac{g}{6}\phi_x^3  \right].
\end{equation*}
(We use the notation $\phi_x=\partial\phi/\partial x$, $\phi_t=\partial\phi/\partial t$, etc.). Setting $\hbar=1$, the Heisenberg equations of motion are 
\begin{equation*}\label{eq:HbergEoM}
	\phi_t = i[H,\phi] = -c\phi_x-\frac{g}{2}\phi_x^2.
\end{equation*}
Introducing $v\equiv g\phi_x$ gives 
\begin{equation*}\label{eq:Hopf}
	v_t+cv_x+v v_x  = 0.
\end{equation*}
The second term is removed by passing to the moving coordinate $x\to x+ct$, in terms of which $v$ obeys the inviscid Burgers equation
\begin{equation}\label{eq:HopfMove}
	v_t+v v_x = 0.
\end{equation}
Classical solutions of Eq.~\eqref{eq:HopfMove} become multivalued when regions of higher velocity $v$ overtake slower regions. In fluid dynamics, this pathology is remedied by the inclusion of dispersion or dissipation, which gives rise to higher gradient terms. At zero temperature there is of course no dissipation, so we add dispersive terms to the  phonon energy. In the moving frame -- so that the linear term is absent -- this now takes the form
\begin{equation*}\label{eq:PhononDispersion}
	\epsilon(k)=-\alpha k^2-\beta k^3+\cdots.
\end{equation*}
The long-wavelength phonon Hamiltonian becomes
\begin{equation}\label{eq:modH}
	H= \int_{-\infty}^\infty dx\left[\frac{\alpha}{2} \phi_x\cH\phi_{xx}-\frac{\beta}{2}\phi_{xx}^2+ \frac{g}{6}\phi_x^3 \right].
\end{equation}
Here $\cH$ denotes the Hilbert transform
\begin{equation*}
	\label{OneLoopRG_Hilbert}
	\cH\phi(x) = \frac{1}{\pi} \int_{-\infty}^\infty \text{P}\frac{\phi(y)}{y-x}dy,
\end{equation*}
where $\text{P}$ indicates the Cauchy principal value. The equation of motion of the Hamiltonian Eq.~\eqref{eq:modH} is
\begin{equation*}\label{eq:NewEoM}
	v_t+v v_x +\alpha \cH v_{xx}+\beta v_{xxx}=0.
\end{equation*}
For $\alpha=0$ we have the Korteweg--de Vries equation, while the case $\beta=0$ corresponds to the Benjamin--Ono (BO) equation~\cite{Benjamin:1967,Ono:1975}. Both equations are completely integrable~\cite{Fokas:1983,Coifman:1990,Kaup:1998}, though the intermediate case is not.

In the following we restrict ourselves to $\beta=0$, though our 	s are applicable to the general case. The resulting Hamiltonian, which we denote $H_{\text{BO}}$, appears as the bosonized description of the Calogero--Sutherland (CS) model of particles of mass $m$ interacting with an inverse square potential $U(x-y)=\frac{\lambda(\lambda-1)}{m(x-y)^2}$ \cite{Minahan:1994a,Polychronakos:1995,Awata:1995}. In this case the coefficients are
\begin{equation*}\label{eq:CSparam}
	g=\frac{\sqrt{2\pi\lambda}}{m},\qquad \alpha=\frac{\lambda-1}{2m}. 
\end{equation*}
$H_\text{BO}$ has also recently been advanced as the effective Hamiltonian (in the moving frame) of edge excitations in the fractional Quantum Hall Effect \cite{Wiegmann:2012}, describing the effects of nonlinearity and dispersion beyond the usual chiral Luttinger liquid theory.

Classically, one of the most dramatic consequences of dispersion is the existence of solitons. For the BO equation these have the particularly simple form $v(x,t)=V(x-v_S t)$, parametrized by the soliton velocity $v_S$ which has the same sign as $\alpha$
\begin{equation}\label{eq:1Soliton}
	V(x) = \frac{4\alpha^2v_S}{v_S^2x^2+\alpha^2}.
\end{equation}
%
Evaluating the energy and momentum 
\[P=\sum_{k>0} k\adop_k\aop_k = \frac{1}{2}\int dx\, \phi_x^2\]
of the soliton gives the dispersion relation $E(P)=(g^2/8\pi \alpha) P^2$. Thus phonons and solitons have opposite dispersion, and in fact correspond to the states of maximum and minimum energy at given momentum. 


The calculation we now describe shows that the dynamical structure factor $S(q,\omega)$ has support only between these two thresholds, with power-law singularities in the vicinity of the edges, given for small $g/\alpha$ by (see Fig.~\ref{fig:Struct})
\begin{equation}\label{eq:Sbehaviour}
	S(q,\omega)\propto\begin{cases}
		\left[\omega-\epsilon(q)\right]^{-1+g^2/8\pi\alpha^2} & \text{for } \omega\gtrsim \epsilon(q)\\
		\left[E(q)-\omega\right]^{-1+8\pi\alpha^2/g^2} & \text{for } \omega\lesssim E(q)
	\end{cases}
\end{equation}
This is consistent with the known exact results for the CS model \cite{Haldane1994,Ha:1994,Minahan:1994a,Lesage:1995,Pustilnik:2006a}. These earlier calculations rely on the complex machinery of Jack symmetric polynomials, which belies the simplicity of the result Eq.~\eqref{eq:Sbehaviour}. Though our calculations are performed in the limit where dispersion dominates the nonlinearity, the form of the result shows that this limit is nontrivial. This is because the nonlinearity is a \emph{marginal} perturbation with respect to the BO dispersion in the sense of the renormalization group, and therefore a resummation of logarithmic divergences is expected. 

\emph{Phonon threshold}. The dynamical structure factor is the Fourier transform of the phonon correlator $\langle v(x,t)v(0,0) \rangle$
\begin{equation}\label{eq:Sdef}
	S(q,\omega) \propto q\int_{-\infty}^\infty \braket{0|\aop_q(t)\adop_q(0)|0}e^{i\omega t}dt,
\end{equation}
where the overall normalization can be fixed by the f-sum rule in a Galilean invariant system. If the phonons are free, i.e. $g=0$, we have
\begin{equation}\label{eq:FreeBose}
	\braket{0|\aop_q(t)\adop_q(0)|0}=e^{-i\epsilon(q) t},
\end{equation}
and $S(q,\omega)$ consists of a single $\delta$-function centred at $\omega=\epsilon(q)$. Now when $g/\alpha$ is nonzero but small, we can expect that for energies and momenta close to the phonon dispersion, the states contributing to $S(q,\omega)$ resemble those of a single phonon. We thus seek a unitary transformation of $H_\text{BO}\to U H_\text{BO}U^\dagger$ to remove the coupling between phonons at leading order. Writing $U = e^A$ in terms of some antihermitian generator gives the condition
\begin{equation*}\label{eq:GenEq}
	\left[A,H_2\right]+H_3=0,
\end{equation*}
with solution $A = \sum_{\{k_i>0\}} A_{k_{1}k_{2}k_{3}}\left(\adop_{k_{1}}\aop_{k_{2}}\aop_{k_{3}}-\text{h.c.}\right)$, where
\begin{equation}
	\label{eq:Acoeff}
	A_{k_{1}k_{2}k_{3}}=\frac{g}{2}\sqrt{\frac{1}{L}}\frac{\sqrt{k_{1}k_{2}k_{3}}}{\alpha k_{1}^{2}-\alpha k_{2}^{2}-\alpha k_{3}^{2}}\delta_{k_1,k_2+k_3}.		
\end{equation}
In considering the effect of the above unitary transformation on a phonon of wavevector $q$, we note that the generator Eq.~\eqref{eq:Acoeff} diverges when one of $k_2$ or $k_3$ approaches zero. This indicates that the phonon has singular interactions with soft phonons that change its momentum very little. Isolating the part of the generator involving one phonon operator with momentum below some small cutoff $\Lambda$, and the others far above gives
\begin{equation*}\label{eq:Asoft}
	\begin{split}
			A_\Lambda &\sim \frac{g}{2\alpha} \sum_{q\gg \Lambda\atop 0<k<\Lambda} \frac{1}{\sqrt{kL}} \left(\adop_q\aop_{q-k}\aop_{k}-\text{h.c.} \right)\\
			&\sim i\frac{g}{2\alpha}\int dx\, \phi_<(x)\rho_>(x).
	\end{split}
\end{equation*}
In the second line, $\phi_<(x)$ indicates the part of the chiral boson involving only $k<\Lambda$, and $\rho_>(x)=\adop_>(x)\aop_>(x)$ is the density of `hard' phonons, where 
\begin{equation*}
\aop_>(x)=\sum_{k\gg \Lambda} a_k e^{ikx}.
\end{equation*}
Performing the unitary transformation generated by $A_\Lambda$ on the hard phonons gives
\begin{equation}\label{eq:aTrans}
		\aop_>(x)\to U^{\vphantom{\dagger}}_\Lambda \aop_>(x)U^\dagger_\Lambda = \exp\left[-i(g/2\alpha)\phi_<(x)\right]\aop_>(x).
\end{equation}
Treating the transformed variables and vacuum as free gives the following approximation to the hard phonon correlation function
\begin{multline*}\label{eq:PhCorr}
			\langle \aop_>(x,t)\adop_>(0,0) \rangle \sim \langle \aop_>(x,t)\adop_>(0,0) \rangle_{H_2}\\
			\times \overbrace{\langle\exp\left[-i(g/2\alpha)\phi_<(x,t)\right]\exp\left[i(g/2\alpha)\phi_<(0,0)\right]\rangle_{H_2}}^{\equiv \cV(x,t)}.
\end{multline*}
Together with Eq.~\eqref{eq:FreeBose} for the free phonon correlation function this gives for Eq.~\eqref{eq:Sdef}
\begin{equation}\label{eq:Convol}
	S(q,\omega) = f(q)\sum_{q'}\tilde\cV(q-q',\omega-\epsilon(q')),
\end{equation}
where $\tilde\cV(q,\omega)$ is the Fourier transform of $\cV(x,t)$, facilitated by splitting $\phi$ into positive ($\phi^+$) and negative ($\phi^-$) wavevectors. 
\begin{equation*}\label{eq:Veval}
	\begin{split}
          \cV(x,t) &\propto \langle \exp \left[-i(g/2\alpha)\phi_<^+(x,t)\right] \exp \left[i(g/2\alpha)\phi_<^-(0,0) \right]\rangle_{H_2}\\
          &= \exp \left[ \frac{g^2}{4\alpha^2} [\phi_<^+(x,t), \phi_<^-(0,0)] \right]\\
          &= \exp \left[\frac{g^2}{8\pi \alpha^2} \int_{1/L}^\Lambda \frac{dk}{k}e^{ikx+i\alpha k^2t}  \right]\\
          &\sim |x|^{-g^2/8\pi \alpha^2},\qquad x^2\gg \alpha t.
	\end{split}
\end{equation*}
Substituting into Eq.~\eqref{eq:Convol} yields the first of Eq.~\eqref{eq:Sbehaviour}.

Let us describe the physical picture underlying this calculation. The hard phonon maintains its identity during interaction with the soft excitations, so may be regarded as a moving impurity. Eq.~\eqref{eq:aTrans} shows that the creation of a hard phonon is associated with a `shake up' of the soft phonon system, as in the orthogonality catastrophe or Fermi edge singularity \cite{Anderson:1967,Schotte:1969}, leading to power law behavior in the vicinity of the phonon threshold. 


\emph{Soliton threshold}. To understand the behaviour in the vicinity of the soliton dispersion, we note that in the large dispersion limit the soliton is \emph{heavy}, (this corresponds to the large repulsion limit of the CS model) which suggests a semiclassical description. 
This is most conveniently implemented within a coherent state functional integral representation of the phonon correlator, which takes the form \cite{Faddeev:1976}
\begin{equation}\label{eq:CorrFnPathInt}
	 q\braket{0|\aop_q(t)\adop_q(0)|0} \propto q\int  \cD\varphi\, \exp\left(iS[\varphi]\right)  \alpha_q(t) \bar\alpha_q(0).
\end{equation}
%
%
$\alpha_q(t)$ is the analog of $a_q(t)$ for the c-number field $\varphi(x,t)$. The action $S[\varphi]=S_\text{BO}[\varphi]+S_\text{B}[\varphi]$ consists of the BO action $S_\text{BO}[\varphi]$, as well as a boundary term $S_\text{B}[\varphi]$ that plays a vital role in the following.
\begin{equation*}\label{eq:Actions}
	\begin{split}
		S_{\text{BO}}[\varphi]&=-\frac{1}{2} \int_0^t d\tau \int dx  \left[\varphi_x\varphi_\tau+\alpha\varphi_x\cH\varphi_{xx}+\frac{g}{3}\varphi_x^3  \right]\\
		S_B[\varphi]&=\frac{1}{2}\int dx\left[\varphi^-\varphi_x^+|_{\tau=0} + \varphi^-\varphi_x^+|_{\tau=t}\right].
	\end{split}
\end{equation*}
We have split the chiral boson into positive and negative wavevector parts $\varphi(x)=\varphi^+(x)+\varphi^-(x)$ analytic in the upper and lower half planes of $x$ respectively.

To implement the semiclassical approximation we consider field configurations close to the soliton: $\varphi(x,\tau)=\Phi(x;X(\tau),\bar X(\tau))+\tilde\varphi(x,\tau)$, where (c.f. Eq.~\eqref{eq:1Soliton})
\begin{equation*}\label{eq:PhiSoliton}
	\Phi(x;X(\tau),\bar X(\tau))=-\frac{2i\alpha}{g}\ln \left(\frac{x- X(\tau)}{x-\bar X(\tau)} \right),
\end{equation*}
with the collective coordinates $X(\tau)$, $\bar X(\tau)$  assumed to be close to a soliton trajectory $v_S \tau\pm i\alpha/v_S$.

%
%

The semiclassical approximation to the correlator then has the form (up to constant factors)
\begin{equation}\label{eq:Gaussian}
\int \cD X\cD\bar X e^{iq[X(0)-\bar X(t)]+iS[\Phi]} \int \cD \tilde\varphi\, e^{i\delta S_\text{B}[\tilde\varphi]+\frac{i}{2}\delta^2S[\tilde\varphi] },
\end{equation}
where the factor $ e^{iq[X(0)-\bar X(t)]}$ originates from the Fourier components of the soliton, and $\delta S_\text{B}[\tilde\varphi]$ arises from the variation of the endpoints
\begin{multline}\label{eq:BoundaryExp}
  \delta S_\text{B}=\int\dd{x} \left[\Phi_x^+\tilde{\varphi}^-|_{\tau=0} - \Phi_x^-\tilde{\varphi}^+|_{\tau=t} \right]= \\\frac{4\pi \alpha}{g} \left[  \tilde{\varphi}^+(X(t),t)-\tilde{\varphi}^-(\bar{X}(0),0)\right].
\end{multline}
The simple poles of the Benjamin soliton lead to the second line of Eq.\eqref{eq:BoundaryExp}, which is completely determined by the soliton `charge'. Even for models without this luxury, at long times any soliton will behave like a delta function in the integrand. 

The computation of the Gaussian integral in Eq.~\eqref{eq:Gaussian} is facilitated by the use of a basis diagonalizing the quadratic action $\delta^2S[\tilde\varphi]$ \cite{Chen:1980}, in terms of which we may write
\begin{multline}\label{eq:phiExpand}
	\tilde\varphi(x,\tau) = \int_0^\infty \frac{dk}{2\pi} \left[\eta(k,\tau)\psi^+(k,y)+\bar\eta(k,\tau)\psi^-(k,y)\right]
\end{multline}
where $y=x-v_St$, and 
\begin{equation}\label{eq:PsiPlus}
	\psi^+(k,y) = \frac{y-X}{y-\bar X}\left[\frac{1}{i(k+v_S/2\alpha)(y-X)}-1 \right]e^{iky}
\end{equation}
Together with functions corresponding to variation of $X(\tau)$ and $\bar X(\tau)$, this basis is complete and orthonormal \cite{Chen:1980}. Substitution into the Gaussian action in Eq.~\eqref{eq:Gaussian} gives 
\begin{equation}\label{eq:ModeSub}
	\begin{split}
		\delta S_{\text{B}} &= \frac{2 \alpha}{g} \int_0^\infty dk \frac{e^{-k\alpha/v_S}}{1+2\alpha k/v_S}\left[\eta(k,t)-\bar \eta(k,0)\right] \\
		\delta^2 S&= \int_0^t d\tau\int \frac{k dk}{\pi} \bar\eta(k,\tau)\left[i\partial_\tau-\omega(k)\right]\eta(k,\tau) \\
		 &+ i\int \frac{dk}{2\pi} k \left[\bar\eta(k,0)\eta(k,0) +\bar\eta(k,t)\eta(k,t) \right] 
	\end{split}
\end{equation}
%
%
where $\omega(k)=-v_S k-\alpha k^2$. Integrating over $\{\eta(k,\tau),\bar\eta(k,\tau)\}$ is now straightforward and yields a factor in the semiclassical correlator equal to $(v_S t/l_S)^{-8\pi \alpha^2/g^2}$ at long times, where $l_S\equiv\alpha/v_S$ is the size of the soliton. 

It remains to perform the integral over the collective coordinates $\{X(\tau),\bar X(\tau\}$. The exponent $q[X(0)-\bar X(t)]+S[\Phi]$ in Eq.~\eqref{eq:Gaussian} is stationary when the collective coordinates follow a soliton trajectory, and the variation at the endpoints fixes $v_S=(g^2/4\pi\alpha)q$ and $E_S=(g^2/8\pi \alpha)q^2$. The Gaussian path integral coincides with that representing the expectation of a free particle propagator in an eigenstate of momentum $q$, and so simply yields a factor $e^{-iE_S t}$.

Combining these elements yields the final expression for the semiclassical structure factor at long times
\begin{equation}\label{eq:SemiclassicalFinal}
	 q\braket{0|\aop_q(t)\adop_q(0)|0} \propto \left(\frac{l_S}{v_S t} \right)^{8\pi\alpha^2/g^2}  \exp(-iE_S t).
\end{equation}
Fourier transformation with respect to time then yields the second of Eq.~\eqref{eq:Sbehaviour}.

In this calculation the soliton edge singularity arises from the linear coupling in $\delta S_\text{B}$ between the soliton and the `phonon' modes parameterized by the $\eta$-variables. The mechanism is then nearly identical to that giving rise to the phonon singularity in our earlier calculation, albeit with inverse coupling, and illustrates the duality between the phonon and soliton pictures. A similar but more heuristic calculation of the absorption threshold due to the creation of dark solitons in a repulsive 1D Bose gas appeared in Ref.~\cite{Khodas:2008}.

In summary, we have shown that, contrary to the prevailing wisdom, nonlinear Quantum Hydrodynamics in one dimension is a tractable quantum field theory. The addition of phonon dispersion allows us to describe the physics of a nonlinear Luttinger liquid. Although our calculation made no explicit use of integrability, the Benjamin--Ono Hamiltonian is integrable at the quantum as well as the classical level \cite{Nazarov:2013}, and it would be interesting to understand the quantum analogs of the classical solitons in more detail.

\bibliography{../Literature/QHD.bib}

\end{document}